\documentclass[aps,showpacs]{revtex4}
\usepackage{psfig}

\def\ep{\epsilon}
\newcommand{\ba}{\begin{eqnarray}}
\newcommand{\ea}{\end{eqnarray}}
\newcommand{\be}{\begin{equation}}
\newcommand{\ee}{\end{equation}}
\newcommand{\mm}{\omega}
\newcommand{\order}[1]{{\cal O}\left(#1\right)}

\begin{document}

\title{
%
%
%

\hfill\hbox{\rm \normalsize Alberta Thy 14-04} \\
\hfill\hbox{\rm \normalsize UVIC--TH--04/08}
\vskip 10pt

Next-to-next-to-leading order calculations for heavy-to-light
decays
}

\author{Ian Blokland$^{1,2}$,
        Andrzej Czarnecki$^{1}$,
        Maciej \'Slusarczyk$^{1,3}$,
        and Fyodor Tkachov$^{4}$}

\affiliation{
1) Department of Physics, University of Alberta, Edmonton, AB T6G 2J1, Canada \\
2) Department of Physics and Astronomy, University of Victoria, Victoria, BC V8P 5C2, Canada \\
3) Institute of Physics, Jagiellonian University, Reymonta 4, 30-059 Krak\'ow, Poland \\
4) Institute for Nuclear Research, Russian Academy of Sciences, Moscow, 117312, Russian Federation\\
}

\begin{abstract}
We present technical aspects of next-to-next-to-leading order
calculations for heavy-to-light decays such as top quark
decay, semileptonic $b$ quark decay into a $u$ quark, muon
decay, and radiative decays like $b\to s\gamma$. Algebraic
reduction of integrals to a set of master integrals is described,
methods of determining the master integrals are presented, and a complete
list of master integrals is given.  As a sample application, the
top quark decay width is calculated to $\order{\alpha_s^2}$ accuracy.
\end{abstract}


\maketitle

\section{Introduction}
This paper describes the technical aspects of the calculation of
$\order{\alpha_s^2}$ (next-to-next-to-leading order, or NNLO)
heavy-to-light quark decays, such as $t\to bW$ and $b\to u l \nu$,
that were presented in~\cite{Blokland:2004ye}.

Computations in the NNLO for massive charged particle decays are
challenging because of the presence of massive propagators which
complicate loop and phase space diagrams.  Here we focus on
processes in which we have one gluon-radiating particle, $Q$, in
the initial state, and one, $q$, in the final state, such as a
semileptonic decay of a quark, $Q\to q + \textrm{leptons}$.
Radiative decays like $b\to s \gamma$ and $b\to s +
\textrm{leptons}$ also fall into this category.  An analogous
process with radiated gluons replaced by photons is the muon decay
$\mu\to e \nu\bar\nu$.

\begin{figure}[htb]
\hspace*{0mm}\psfig{figure=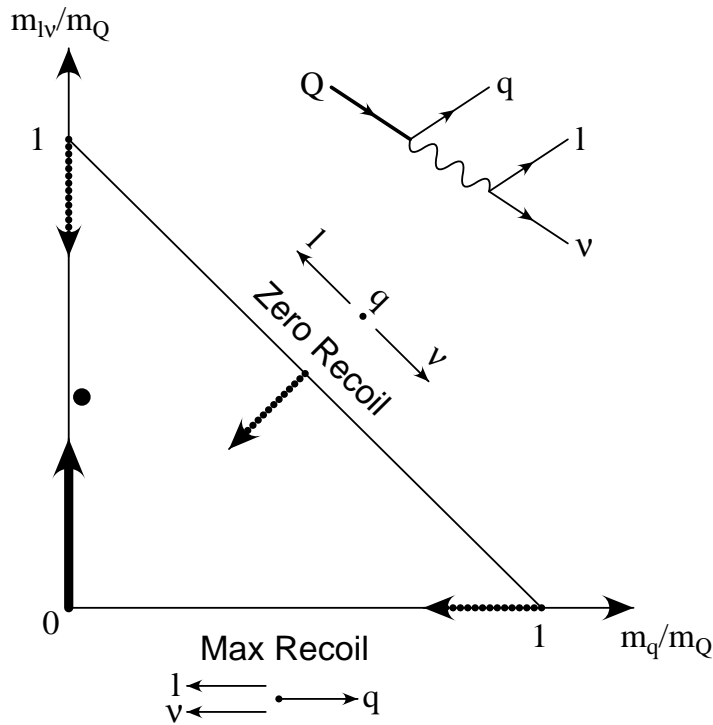,width=100mm}
\caption{Kinematic boundaries of the semileptonic decays $Q\to
q+ \mbox{leptons}$. The solid arrow shows the expansion presented
in this paper.  Previously known expansions are indicated with
dotted arrows
\protect\cite{Czarnecki:1997hc,Czarnecki:1998kt,Czarnecki:2001cz}.
Analytical results are now known along the whole zero-recoil line
\protect\cite{Czarnecki:1997cf,Franzkowski:1997vg} and at the
lower left vertex of the triangle (this work).  The dot at the
left of the triangle indicates the kinematic configuration
corresponding to the decay $t\to bW$.} \label{fig:range}
\end{figure}

A useful approach to such problems consists in expanding the decay
amplitudes around some limit in which the diagrams can be computed
analytically.  In the past, all such expansions started with some
point on the so-called zero-recoil line (see
Fig.~\ref{fig:range}).  Zero-recoil refers to the kinematic case
in which the produced charged particle $q$ remains at rest in the
rest frame of the decaying $Q$.

In the present project we address the other kinematic extreme
wherein a massive $Q$ decays into two massless particles.  One of
the latter may be a pair of collinear massless leptons or a $W$
boson in the limit of $Q$ much heavier than the $W$.  For a
physical top quark decay, we can expand in $m_W/m_t$. Such an
expansion, if taken to a sufficiently high order, also helps to
describe the lepton invariant mass spectrum in the semileptonic
$b\to u $ decay.

The main technical difficulty in constructing this expansion is
the evaluation of a class of diagrams in which the four-momenta of
all the final-state particles are on the order of the mass of the
decaying quark.  We explain here details of the methods we have
used to overcome this obstacle.

This paper is organized as follows.  In Section II we describe the
calculational framework: the optical theorem, the various
contributions needed to construct the expansion, and the ways of
computing the necessary integrals.  In Section III, we list the
topologies of the Feynman integrals and give a set of so-called
primitive or master integrals to which all others can be
algebraically reduced.  In Section IV, we present explicit
examples of how typical primitive integrals can be evaluated.
Section V contains our results and Section VI summarizes the
paper.  Throughout the presentation, we use top quark decay to
illustrate the technical steps.

\section{Calculational Framework}
\subsection{Optical theorem}
At $\order{\alpha_s^2}$, there are three classes of radiative
corrections which contribute to the decay $t\rightarrow bW$:
diagrams with virtual gluon loops, diagrams with real radiation,
and diagrams with both a virtual gluon loop and a real gluon
emission.  The phase space calculations that would be required to
calculate the decay rate in this approach are difficult even in
the case that both masses $m_b$ and $m_W$ are neglected, and as we
mentioned in the introduction, finding the functional dependence of the
decay rate on $m_W$ is one of our objectives.
Our task becomes manageable by using the optical theorem
to relate the decay width to the imaginary part of top quark
self-energy diagrams:
\be
\Gamma = \frac{\mathrm{Im}(\Sigma)}{m_t}
\ . \label{optthmtop}
\ee
The self-energy diagrams, $\Sigma$, that
we must consider for the $\order{\alpha_s^2}$ decay rate are
three-loop diagrams (two gluon loops and a $t\to bW \to t$ loop).
The imaginary parts of these diagrams arise from the various cuts
that can be made through the $b$, $W$, and $g$ lines so that the
two sides of the cut correspond to particular decay amplitudes.
Despite the introduction of an additional loop, the calculation is
feasible as a result of an assortment of techniques which have
been developed to solve multiloop integrals analytically with the
help of symbolic computation software.

\subsection{Asymptotic expansions}
With $m_b=0$, there are two scales in the problem: $m_t$ and
$m_W$.  We define an expansion parameter $\mm=m_W^2/m_t^2$ so that
the two scales can be expressed as hard and soft ($\order{1}$ and
$\order{\sqrt{\mm}}$, respectively) using $m_t$ as the unit of
energy.
The asymptotic expansion in powers and logarithms of $\mm$
of the contributing integrals is performed
using the method of asymptotic operation~\cite{AE},
and contributions associated with the two scales are identified
using the following heuristic mnemonic.
The loop momenta flowing through the
gluon lines must be hard or else scaleless integrals (which
vanish in dimensional regularization) will arise.  This leaves us
with two regions to consider, depending on the scale of the loop
momentum flowing through the $W$ line. In the first region, all
the loop momenta are hard and the Euclidean-space $W$ propagator
can be expanded as a series
of massless
propagators:
\be \frac{1}{k^2+m_W^2} = \frac{1}{k^2} -
\frac{m_W^2}{k^4} + \frac{m_W^4}{k^6} - \frac{m_W^6}{k^8} + \ldots
\ . \label{khard}
\ee
After the integration, powers of $m_W^2/k^2$ will become powers of $\mm$.

In the second region, the gluon momenta are
hard but the loop momentum flowing through the $W$ is soft.  As a
result, this loop momentum ($k$) decouples from the other momenta
($\ell$) flowing through the $b$ quark lines via the expansion \be
\frac{1}{(k+\ell)^2} = \frac{1}{\ell^2} -
\frac{(k^2+2k\ell)}{\ell^4} + \frac{(k^2+2k\ell)^2}{\ell^6} -
\frac{(k^2+2k\ell)^3}{\ell^8} + \ldots \ . \label{ksoft} \ee
Consequently, the diagrams in this region factor into a product of
a two-loop self-energy-type integral and a one-loop vacuum bubble
integral with a scale of $m_W$. The leading contribution from this
second region is $\order{\mm^2}$; the interplay between the
two regions gives rise to terms with a formally large logarithm $\ln \mm$.

\subsection{Single-scale integrals}
With the asymptotic expansions in place to extract the functional
dependence of the top decay width on $m_W$, we have reduced the
problem to that of solving single-scale self-energy loop
integrals.  The indispensable tools for doing this are
integration-by-parts identities, used to
reduce loop integrals with arbitrary powers of denominator
factors to a small set of master integrals~\cite{Tkachov:1981wb}.  There are two
approaches with which these identities can be applied to a
collection of loop integrals.  In the traditional method,
one inspects the structure of the
identities and rearranges them manually into the form of
recurrence relations for a complete solution of the
system. This method has proven to be very
successful in numerous applications (e.g., \cite{Larin:1991fx,Larin:1991zw,bro91a,
vanRitbergen:1998yd, vanRitbergen:1999gs, Czarnecki:2001cz}) but
it requires much human work to implement.

Recently, Laporta~\cite{Laporta:2001dd} succeeded in developing a
targeted solution of the linear system
(for only those integrals that are needed in a given problem)~\cite{targeted}.
 This was not previously practical as it is much
more expensive computationally. In our calculation  we
used the traditional complete approach (programmed in FORM~\cite{Vermaseren:2000nd}) as well as a modified version of the new targeted
algorithm for which we implemented a dedicated computer algebra
program. In both cases we independently obtained identical results.
This serves as a check of correctness and also enables us to
compare the two methods~\cite{csc}.

\section{List of Topologies and Master Integrals}
\label{sec:topo}
In the ``soft'' loop-momentum region, the diagrams factor into a
product of a two-loop self-energy-type integrals and a one-loop
vacuum bubble integral with a scale of $m_W$.  The bubble integral
is purely real and can be solved easily.  The entire set of
two-loop self-energy integrals was first solved
in~\cite{bielefeld}; the contributions from this region are
therefore quite straightforward.

\begin{figure}[htb]
\begin{tabular}{ccc}
\psfig{figure=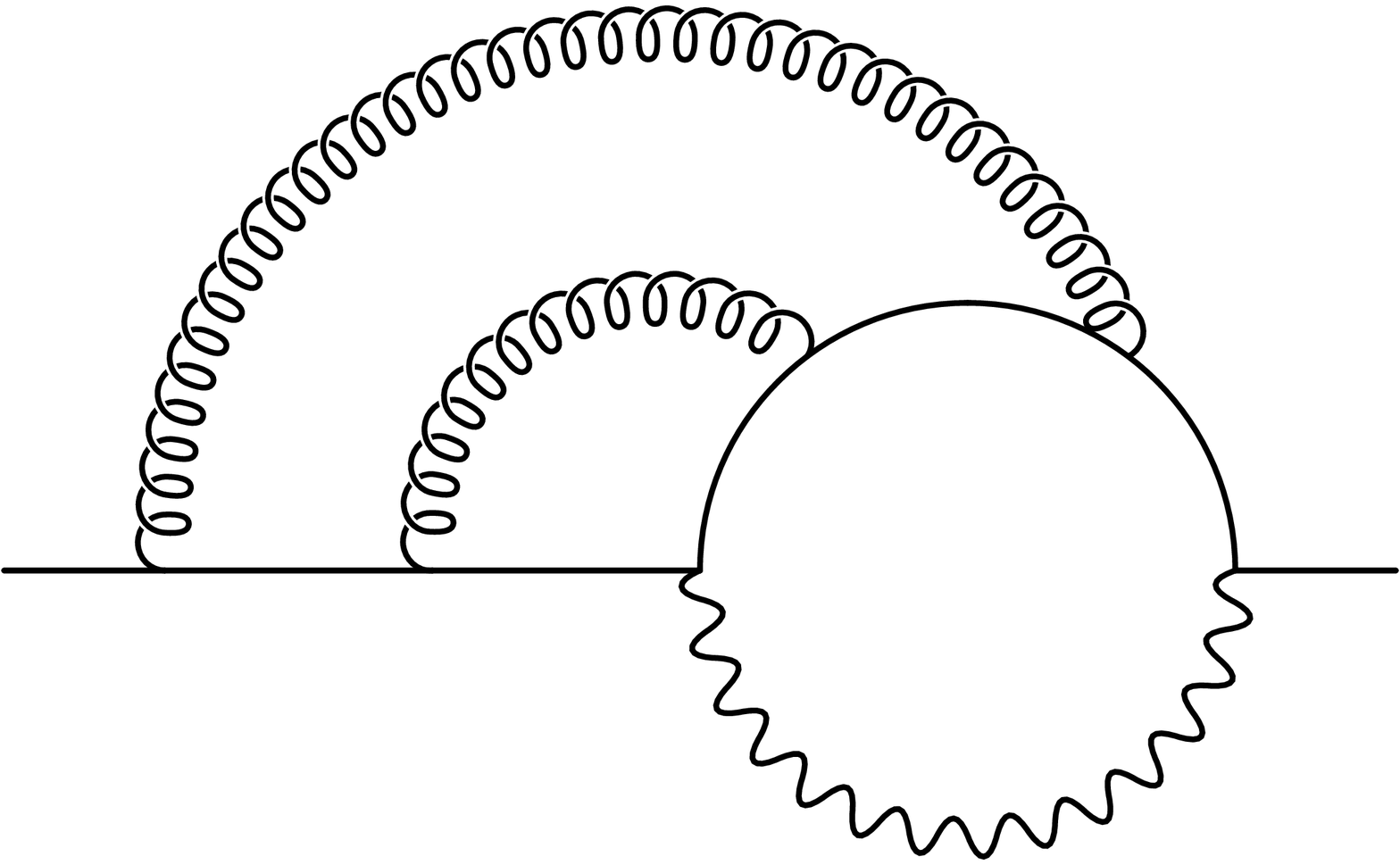,width=52mm}
&
\hspace{5mm}
\psfig{figure=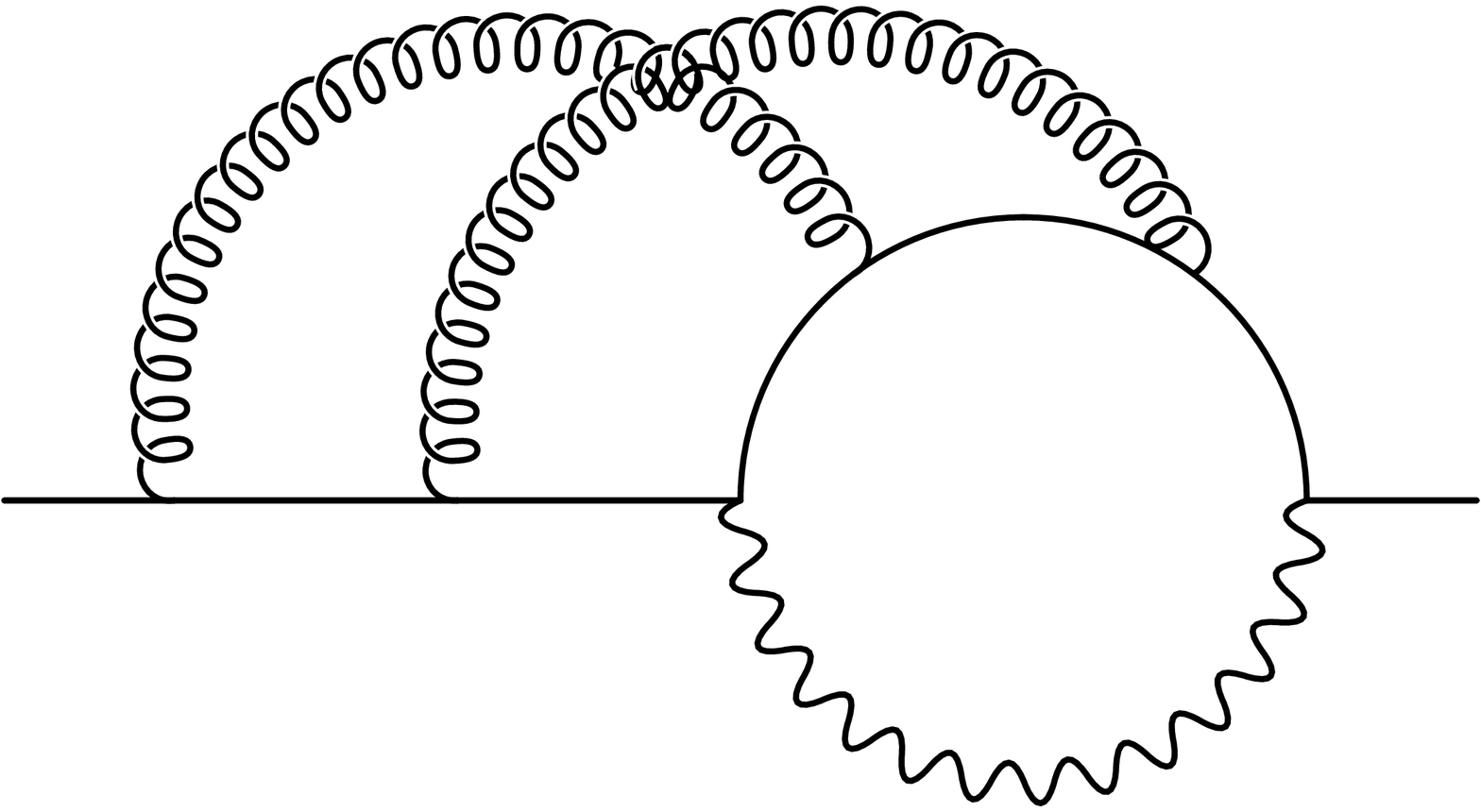,width=52mm}
\hspace{5mm}
&
\psfig{figure=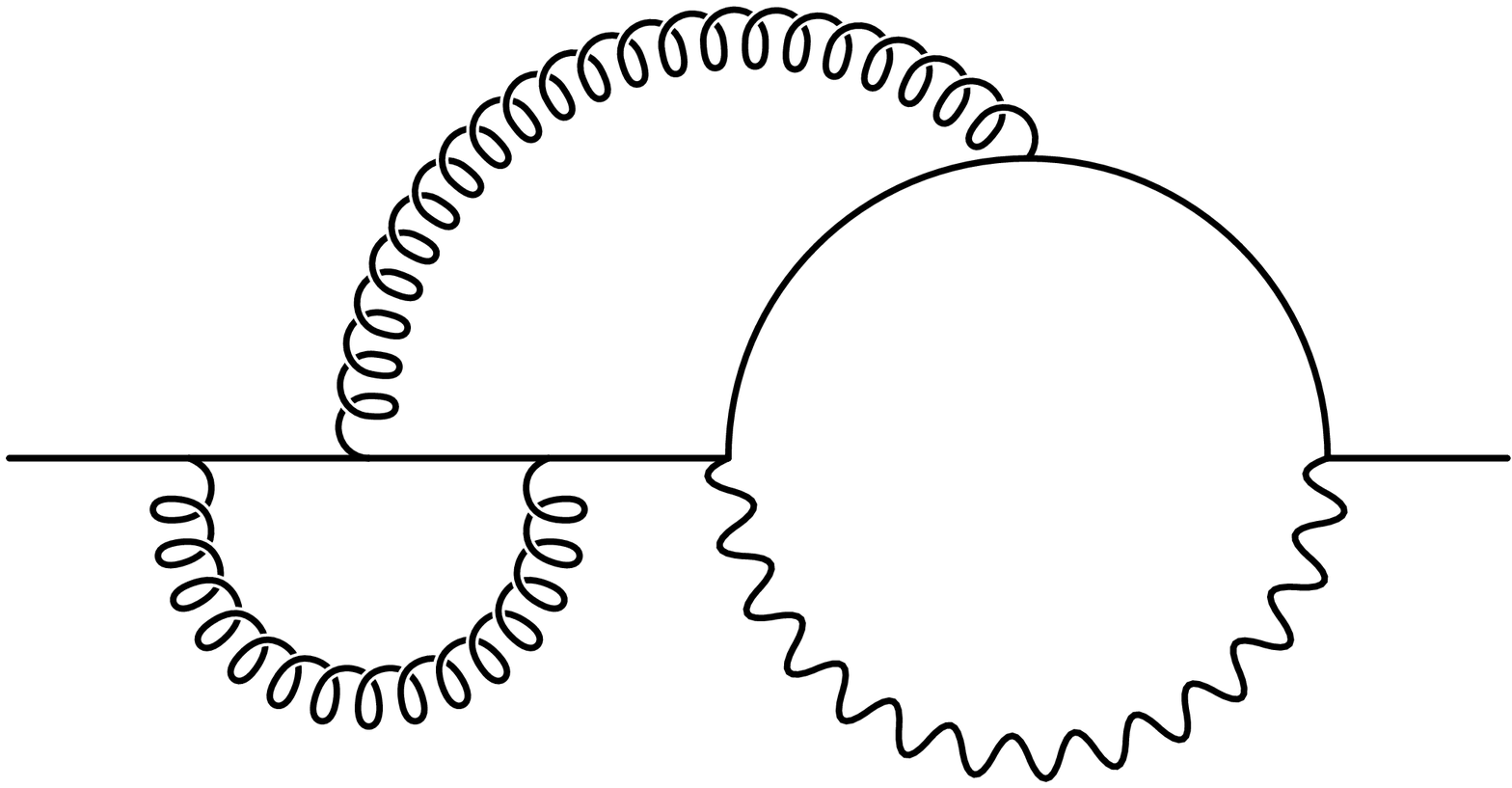,width=50mm}
\\
A & B & C
\\ \\ \\
\psfig{figure=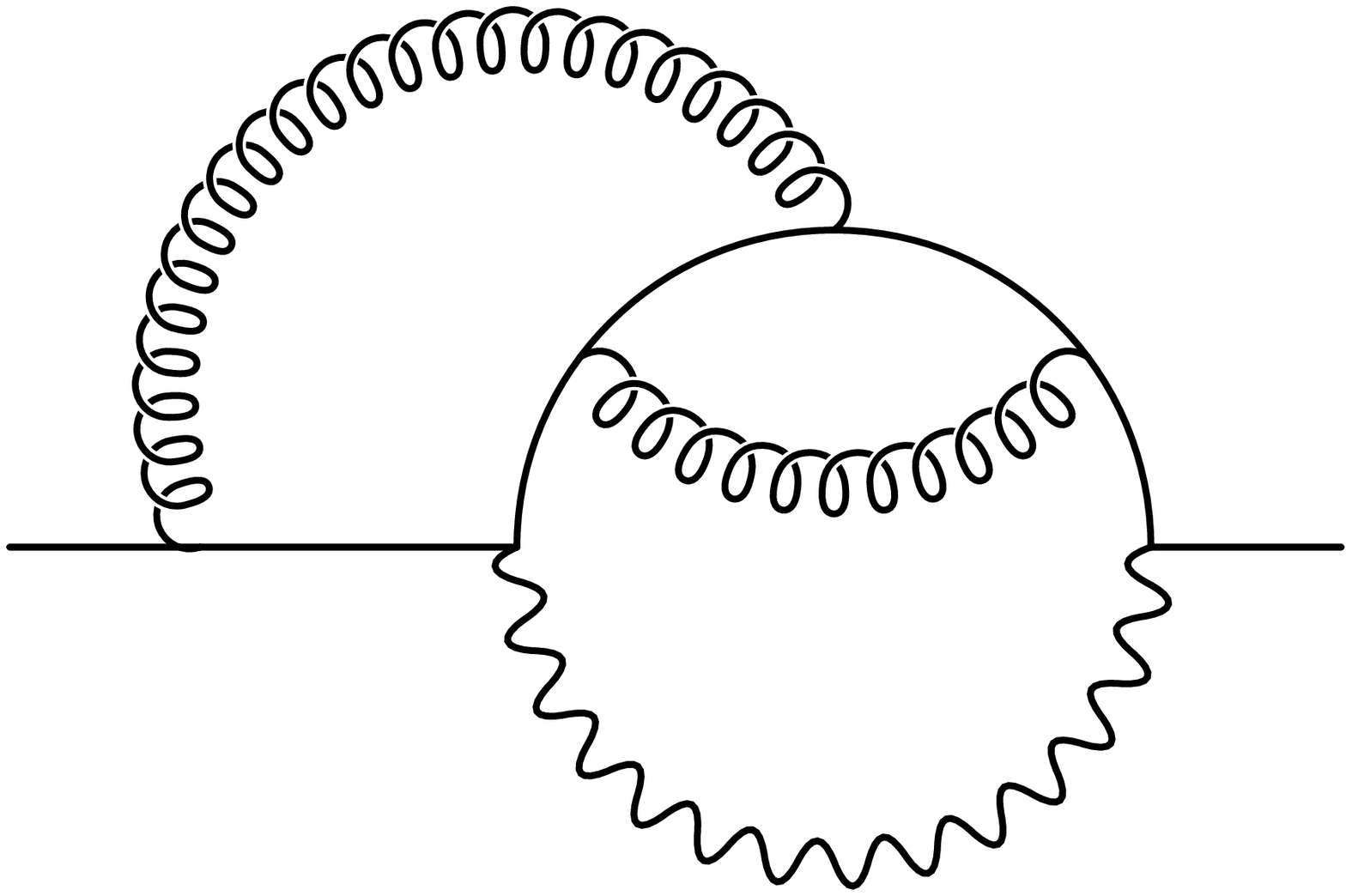,width=42mm}
&
\hspace{5mm}
\psfig{figure=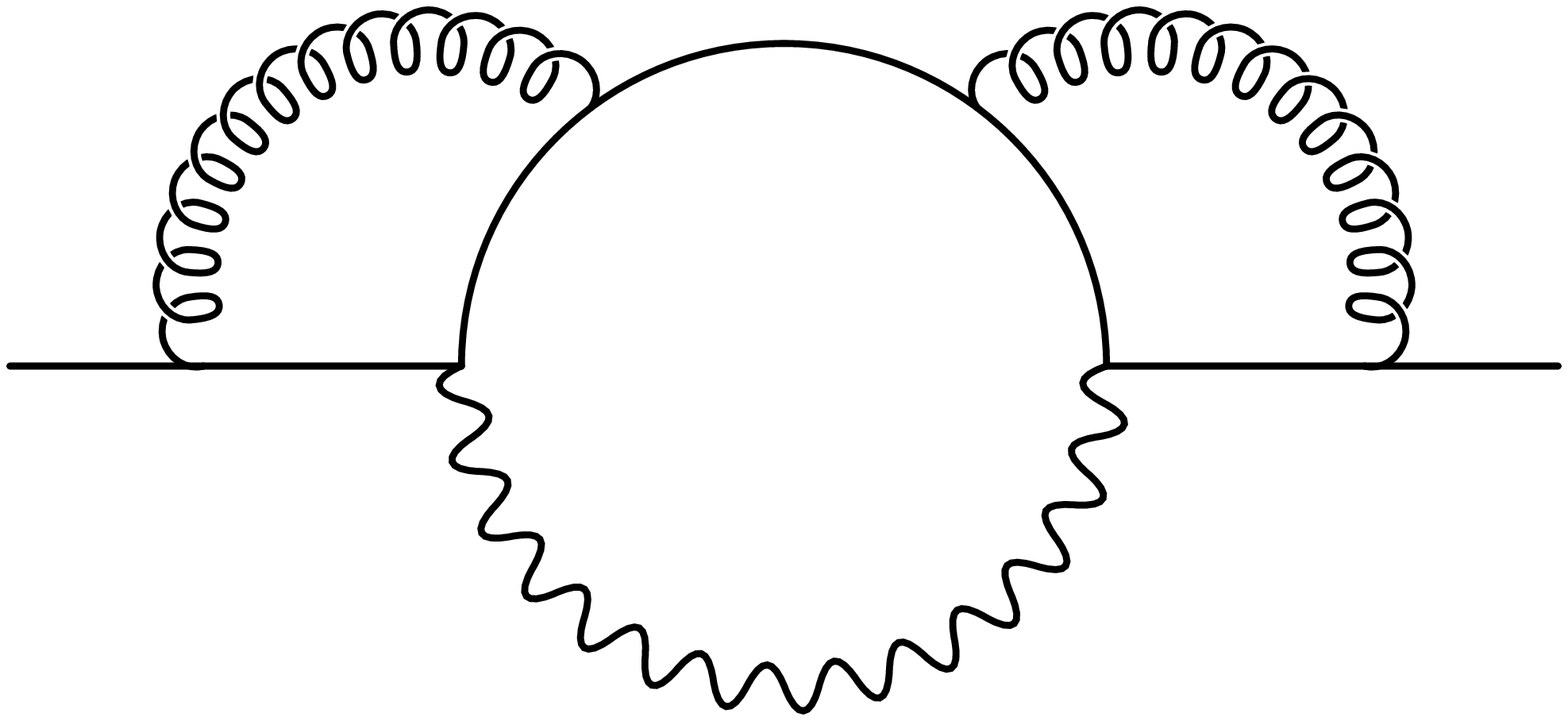,width=48mm}
\hspace{5mm}
&
\psfig{figure=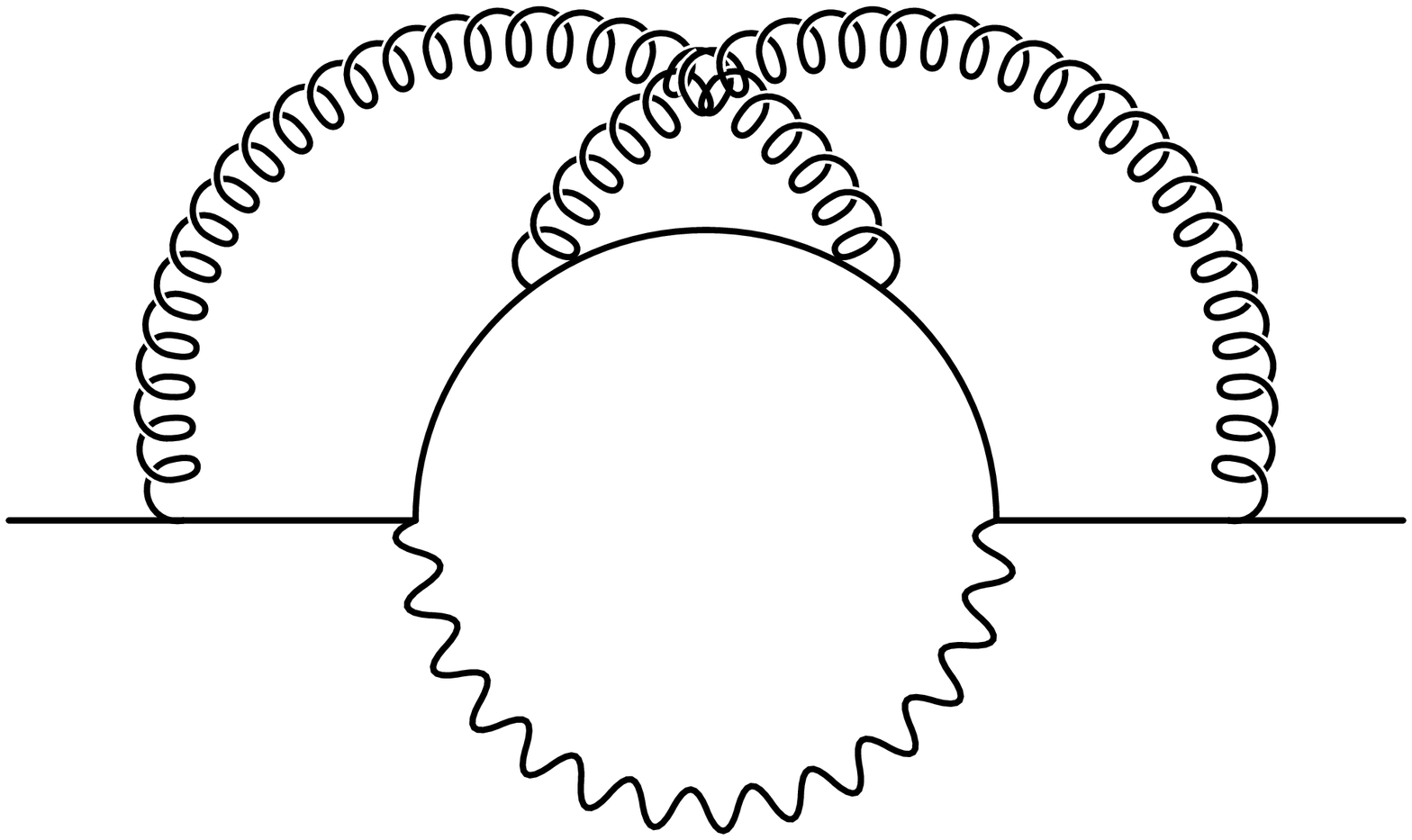,width=48mm}
\\
D & E & F
\\ \\ \\
\psfig{figure=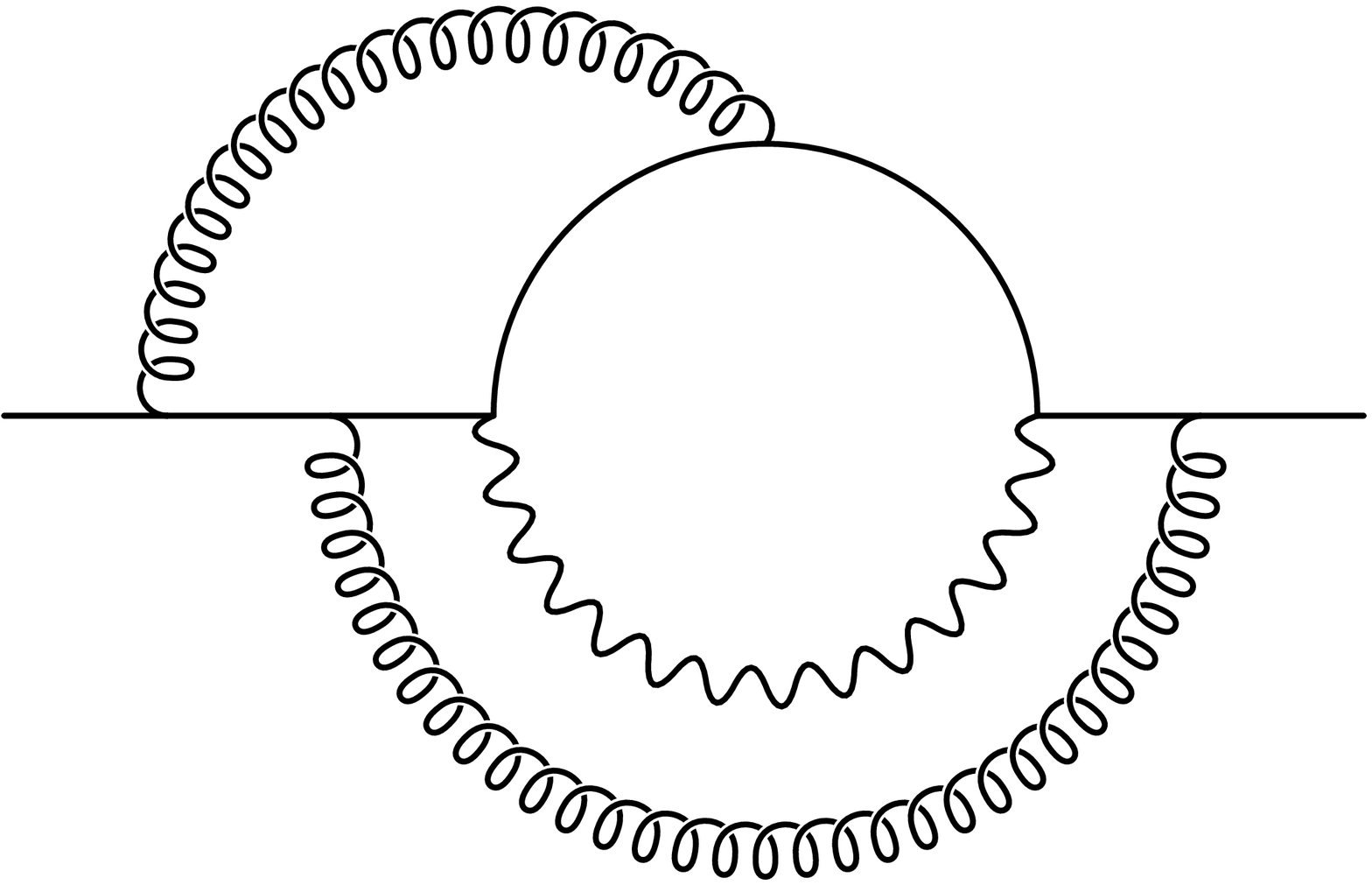,width=50mm}
&
\hspace{5mm}
\psfig{figure=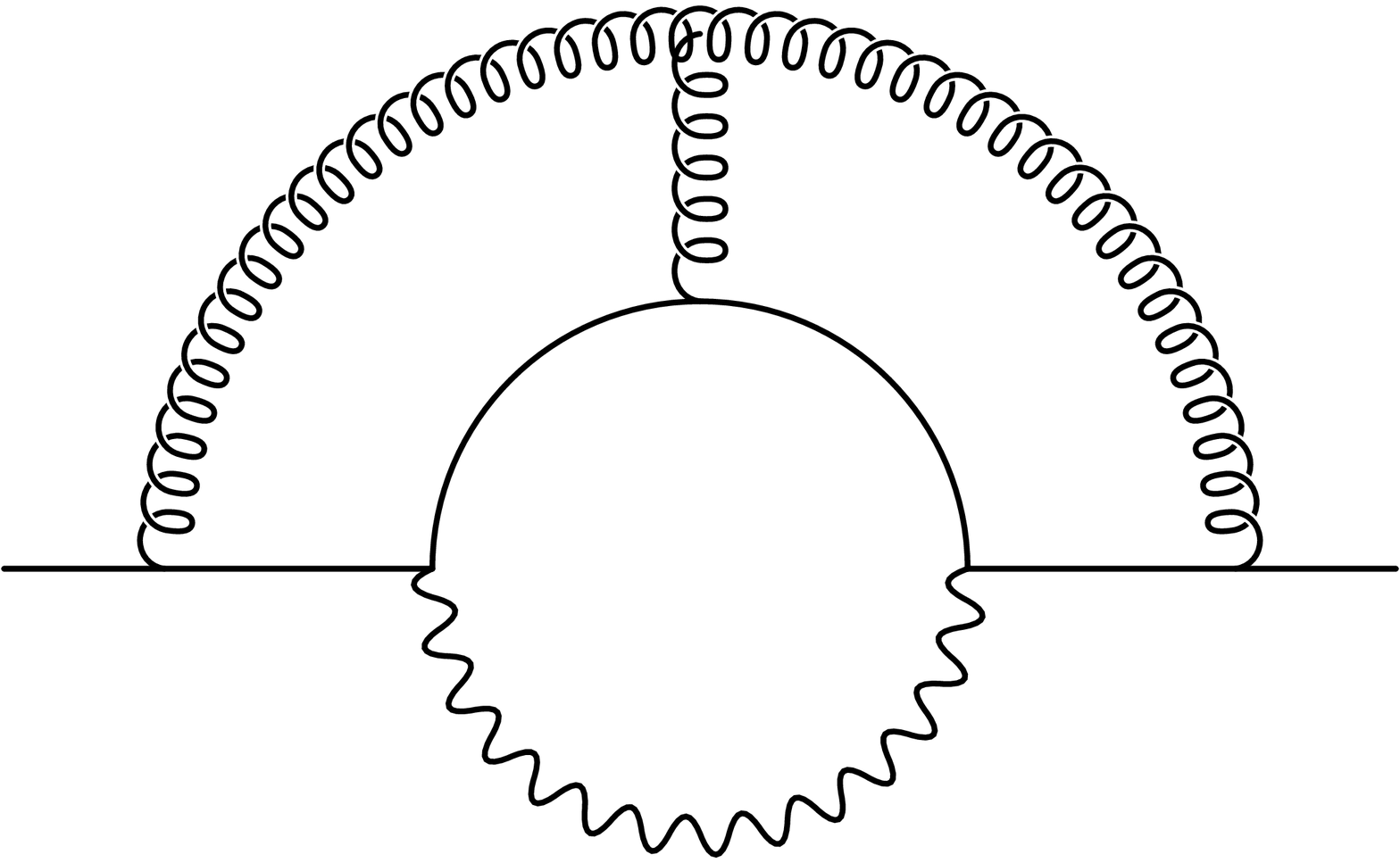,width=52mm}
\hspace{5mm}
&
\psfig{figure=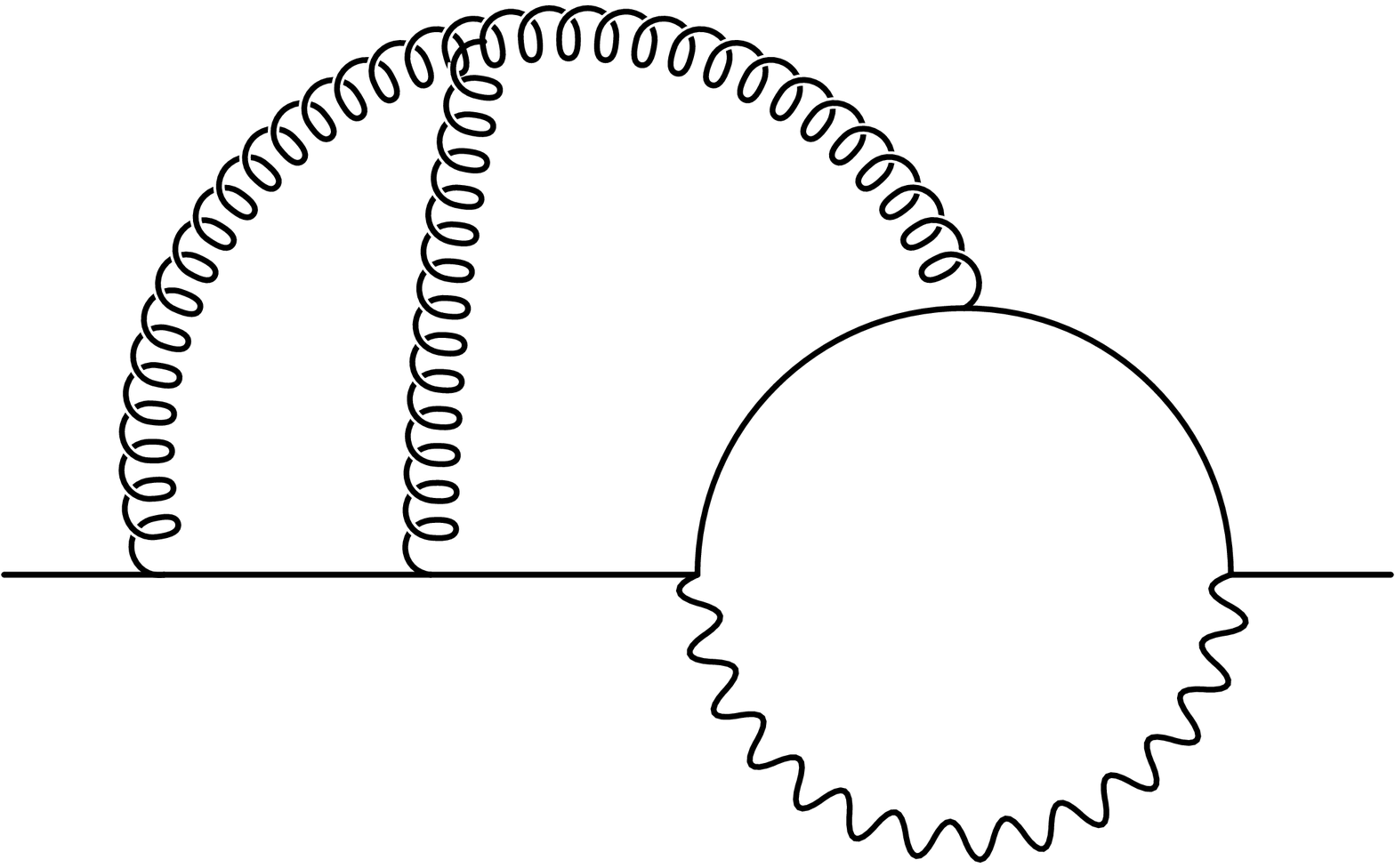,width=52mm}
\\
G & H & I
\end{tabular}
\caption{The nine $\mathcal{O}(\alpha_s^2)$ diagrams
for $t\rightarrow bW$ which constitute the master three-loop topologies of the calculation.}
\label{fig:topologies}
\end{figure}

In the ``hard'' loop-momentum region, we need to extract the
imaginary parts of genuine three-loop integrals.  We identify 9
master topologies in terms of which all 36 $\order{\alpha_s^2}$
diagrams can be expressed.  The diagrams which give rise to these
master topologies are shown in Fig.~\ref{fig:topologies}.  They
are labeled by the letters $A$ through $I$ in analogy with the
notation of~\cite{vanRitbergen:1999fi} for the $\order{\alpha^2}$
contributions to muon decay.
We use dimensional regularization ($D\equiv 4-2\epsilon$) and take the top quark
mass as the unit of energy, $m_t\equiv 1$.  Using the notation $[d^Dk]=d^Dk/(2\pi)^D$
we define the following Euclidean-space loop integrals ($p$ is the incoming momentum),
\ba
A(a_1,a_2,a_3,a_4,a_5,a_6,a_7,a_8,a_9) & = & \int \frac{[d^Dk_1][d^Dk_2][d^Dk_3]}{k_1^{2a_1}(k_1+p)^{2a_2}(k_1+k_3+p)^{2a_3} (k_1+k_2+k_3+p)^{2a_4}}  \nonumber \\
&& \hspace{10mm} \times \ \frac{1}{[(k_2+k_3+p)^2+1]^{a_5}(k_3^2+2k_3p)^{a_6}k_2^{2a_7}k_3^{2a_8}(2k_2p)^{a_9}} \\
B(a_1,a_2,a_3,a_4,a_5,a_6,a_7,a_8,a_9) & = & \int \frac{[d^Dk_1][d^Dk_2][d^Dk_3]}{k_1^{2a_1}(k_1+p)^{2a_2}(k_1+k_2+p)^{2a_3} (k_1+k_2+k_3+p)^{2a_4}}  \nonumber \\
&& \hspace{10mm} \times \ \frac{1}{[(k_2+k_3+p)^2+1]^{a_5}(k_3^2+2k_3p)^{a_6}k_2^{2a_7}k_3^{2a_8}(2k_2p)^{a_9}} \\
C(a_1,a_2,a_3,a_4,a_5,a_6,a_7,a_8,a_9) & = & \int \frac{[d^Dk_1][d^Dk_2][d^Dk_3]}{k_1^{2a_1}(k_1+p)^{2a_2}(k_1+k_2+p)^{2a_3}k_2^{2a_4} (k_2^2+2k_2p)^{a_5}}  \nonumber \\
&& \hspace{10mm} \times \ \frac{1}{[(k_2+k_3+p)^2+1]^{a_6}(k_3^2+2k_3p)^{a_7}k_3^{2a_8}(2k_1k_3)^{a_9}} \\
D(a_1,a_2,a_3,a_4,a_5,a_6,a_7,a_8,a_9) & = & \int \frac{[d^Dk_1][d^Dk_2][d^Dk_3]}{k_1^{2a_1}(k_1+p)^{2a_2}(k_1+k_2+p)^{2a_3} (k_1+k_2+k_3+p)^{2a_4}}  \nonumber \\
&& \hspace{10mm} \times \ \frac{1}{(k_1+k_3+p)^{2a_5}k_2^{2a_6}k_3^{2a_7}(k_3^2+2k_3p)^{a_8}(2k_2p)^{a_9}} \\
E(a_1,a_2,a_3,a_4,a_5,a_6,a_7,a_8,a_9) & = & \int \frac{[d^Dk_1][d^Dk_2][d^Dk_3]}{k_1^{2a_1}(k_1+p)^{2a_2}(k_1+k_2+p)^{2a_3} (k_1+k_3+p)^{2a_4}k_2^{2a_5}}  \nonumber \\
&& \hspace{10mm} \times \ \frac{1}{k_3^{2a_6}(k_2^2+2k_2p)^{a_7}(k_3^2+2k_3p)^{a_8}(2k_2k_3)^{a_9}} \\
F(a_1,a_2,a_3,a_4,a_5,a_6,a_7,a_8,a_9) & = & \int \frac{[d^Dk_1][d^Dk_2][d^Dk_3]}{k_1^{2a_1}(k_1+k_2+k_3+p)^{2a_2}(k_1+k_2+p)^{2a_3} (k_1+k_3+p)^{2a_4}}  \nonumber \\
&& \hspace{10mm} \times \ \frac{1}{k_2^{2a_5}k_3^{2a_6}(k_2^2+2k_2p)^{a_7}(k_3^2+2k_3p)^{a_8}(2k_2 k_3)^{a_9}} \\ 
G(a_1,a_2,a_3,a_4,a_5,a_6,a_7,a_8,a_9) & = & \int \frac{[d^Dk_1][d^Dk_2][d^Dk_3]}{k_1^{2a_1}(k_1+k_2+p)^{2a_2}(k_1+k_2+k_3+p)^{2a_3} (k_2^2+2k_2p)^{a_4}}  \nonumber \\
&& \hspace{10mm} \times \ \frac{1}{[(k_2+k_3+p)^2+1]^{a_5}(k_3^2+2k_3p)^{a_6}k_2^{2a_7}k_3^{2a_8}(2k_1p)^{a_9}} \\
H(a_1,a_2,a_3,a_4,a_5,a_6,a_7,a_8,a_9) & = & \int \frac{[d^Dk_1][d^Dk_2][d^Dk_3]}{k_1^{2a_1}(k_1+k_2+p)^{2a_2}(k_1+k_3+p)^{2a_3} k_2^{2a_4}k_3^{2a_5}(k_2^2+2k_2p)^{a_6}}  \nonumber \\
&& \hspace{10mm} \times \ \frac{1}{(k_3^2+2k_3p)^{a_7}(k_2-k_3)^{2a_8}(2k_1p)^{a_9}} \\
I(a_1,a_2,a_3,a_4,a_5,a_6,a_7,a_8,a_9) & = & \int \frac{[d^Dk_1][d^Dk_2][d^Dk_3]}{k_1^{2a_1}(k_1+p)^{2a_2}(k_1+k_2+p)^{2a_3} (k_2^2+2k_2p)^{a_4}(k_3^2+2k_3p)^{a_5}}  \nonumber \\
&& \hspace{10mm} \times \ \frac{1}{k_2^{2a_6}k_3^{2a_7}(k_2-k_3)^{2a_8}(2k_1k_3)^{a_9}}
\ea
For the decay rate calculation, we need the imaginary parts of these integrals,
obtained by setting $p^2 = -1 + i0$.
For the exponent $a_9$ we only need non-positive values;
it refers to a product of momenta that may remain in the numerator after canceling scalar products
against denominator factors.

In order to specify the boundary conditions of the recurrence relations, we need
24 master integrals.
They vary greatly in their complexity, but for completeness,
we shall list them all.  Some of these integrals --- particularly those with six or fewer lines
--- are used by more than one of the topologies and thus any descriptions we provide using
the integral labels $A$ through $I$ are not necessarily unique.  $\mathcal{F}$ denotes the loop
factor $\Gamma(1+\ep)/(4\pi)^{D/2}$.

Five of the master integrals can be solved in closed form to all orders in $\ep$
via a sequence of simple one-loop integrals:
\ba
\mathrm{Im} \left( \int \frac{[d^Dk_1]}{k_1^2(k_1+p)^2} \int \frac{[d^Dk_2]}{(k_2^2+1)} \int \frac{[d^Dk_3]}{(k_3^2+1)} \right)
& = & \pi \mathcal{F}^3 \left[ \frac{1}{\ep^2} + \frac{4}{\ep} + \left( 11 - \frac{\pi^2}{3} \right) + \left( 26 - \frac{4\pi^2}{3} - 2\zeta_3 \right) \ep \right. \nonumber \\
& & \qquad \left. + \left( 57 - \frac{11\pi^2}{3} - 8\zeta_3 + \frac{\pi^4}{90} \right) \ep^2 + \order{\ep^3} \right] \ , \\
\mathrm{Im} \left( \int \frac{[d^Dk_1][d^Dk_2]}{k_1^2k_2^2(k_1+k_2+p)^2} \int \frac{[d^Dk_3]}{(k_3^2+1)} \right)
& = & \pi \mathcal{F}^3 \left[ -\frac{1}{2\ep}  - \frac{15}{4} + \left( -\frac{145}{8} + \frac{\pi^2}{2} \right) \ep \right. \nonumber \\
& & \qquad + \left( -\frac{1155}{16} + \frac{15\pi^2}{4} + 5\zeta_3 \right) \ep^2 \\
& & \qquad \left. + \left( -\frac{8281}{32} + \frac{145\pi^2}{8} + \frac{75\zeta_3}{2} - \frac{7\pi^4}{60} \right) \ep^3 + \order{\ep^4} \right] \ , \nonumber \\
\mathrm{Im} \left( \int \frac{[d^Dk_1][d^Dk_2][d^Dk_3]}{k_1^2k_2^2k_3^2(k_1+k_2+k_3+p)^2} \right)
& = & \pi \mathcal{F}^3 \left[ \frac{1}{12} + \frac{71}{72}\ep +
 \left( \frac{3115}{432} - \frac{\pi^2}{6} \right) \ep^2 \right. \nonumber \\
& & \qquad + \left( \frac{109403}{2592} - \frac{71\pi^2}{36} - \frac{7\zeta_3}{3} \right) \ep^3
 \\
& & \qquad \left. + \left( \frac{3386467}{15552} - \frac{3115\pi^2}{216} - \frac{497\zeta_3}{18}
 + \frac{4\pi^4}{45} \right) \ep^4 + \order{\ep^5} \right] \ , \nonumber \\
\mathrm{Im} \left( \int \frac{[d^Dk_1][d^Dk_2][d^Dk_3]}{(k_1+p)^2(k_1+k_2)^2(k_1+k_3)^2k_2^2k_3^2} \right)
& = & \pi \mathcal{F}^3 \left[ \frac{1}{\ep} + 10 + \left( 64 - 2\pi^2 \right) \ep + \left( 336 - 20\pi^2 - 22\zeta_3 \right) \ep^2 \right. \nonumber \\
& & \qquad \left. + \left( 1584 - 128\pi^2 - 220\zeta_3 + \frac{7\pi^4}{6} \right) \ep^3 + \order{\ep^4} \right] \ , \\
\mathrm{Im} \left( \int \frac{[d^Dk_1]}{k_1^2(k_1+p)^2}
 \int \frac{[d^Dk_2][d^Dk_3]}{k_2^2k_3^2[(k_1+k_2+p)^2+1]} \right)
& = & \pi \mathcal{F}^3 \left[ -\frac{1}{2\ep^2}  - \frac{9}{4\ep} + \left( -\frac{47}{8} - \frac{\pi^2}{6} \right) + \left( -\frac{133}{16} - \frac{3\pi^2}{4} - 3\zeta_3 \right) \ep \right. \nonumber \\
& & \qquad \left. + \left( \frac{417}{32} - \frac{47\pi^2}{24} - \frac{27\zeta_3}{2} - \frac{\pi^4}{4} \right) \ep^2 + \order{\ep^3} \right] \ .
\ea
Two more master integrals are also a sequence of one-loop integrals,
except that the final one-loop integral is less trivial:
\ba
\mathrm{Im} \left( \int \frac{[d^Dk_1][d^Dk_2][d^Dk_3]}{[(k_1+p)^2+1](k_1+k_2)^2(k_1+k_3+p)^2k_2^2k_3^2} \right)
& = & \pi \mathcal{F}^3 \left[ \left( \frac{7}{4} - \frac{\pi^2}{6} \right) + \left( \frac{175}{8} - \frac{3\pi^2}{4} - 11\zeta_3 \right) \ep \right. \\
& & \qquad \left. + \left( \frac{2681}{16} - \frac{131\pi^2}{24} - \frac{99\zeta_3}{2} - \frac{29\pi^4}{60} \right) \ep^2 + \order{\ep^3} \right] \ , \nonumber \\
\mathrm{Im} \left( \int \frac{[d^Dk_1][d^Dk_2][d^Dk_3]}{k_1^2[(k_1+p)^2+1](k_1+k_2+k_3+p)^2k_2^2k_3^2} \right)
& = & \pi \mathcal{F}^3 \left[ \frac{1}{4} + \frac{25}{8}\ep + \left( \frac{383}{16} - \frac{\pi^2}{2} \right) \ep^2 + \order{\ep^3} \right] \ .
\ea
One of the master integrals requires the real part of the two-loop massive self-energy integral first
evaluated in~\cite{Broadhurst:1991fy}:
\ba
\mathrm{Im} \left( \int \frac{[d^Dk_1]}{k_1^2(k_1+p)^2}
\int \frac{[d^Dk_2][d^Dk_3]}{(k_2^2+1)(k_3^2+1)[(k_2+k_3+p)^2+1]} \right)
& = & \pi \mathcal{F}^3 \left[ -\frac{3}{2\ep^2} - \frac{29}{4\ep}
+ \left( -\frac{175}{8} + \frac{\pi^2}{2} \right) \right. \nonumber \\
& & \qquad + \left( -\frac{765}{16} + \frac{13\pi^2}{12} + 3\zeta_3 \right) \ep \\
& & \hspace{-12mm} \left. + \left( -\frac{1943}{32} -
\frac{97\pi^2}{24} - \frac{27\zeta_3}{2} - \frac{\pi^4}{60} +
8\pi^2\ln2 \right) \ep^2 + \order{\ep^3} \right] \ . \nonumber
\ea
Five of the master integrals are only needed to leading order in
our calculations and are very closely related to some of the
four-loop integrals calculated for muon decay
in~\cite{vanRitbergen:1999fi}.  Specifically, if the massless
neutrino-electron loop is integrated out, an overall factor of
$B(1-\ep,1-\ep)\mathcal{F}/\ep$ is obtained and the exponent on
the $W$ line increases from $1$ to $(1+\ep)$.  To leading order in
$\ep$, neither this modification to the $W$ exponent nor the
difference in the normalization conventions changes the master
integral, and so after converting the appropriate results from
Minkowski space to Euclidean space, we conveniently obtain the
following master integrals: \ba
\mathrm{Im}\ A(1,1,0,1,1,1,1,1,0)
& = & \pi \mathcal{F}^3 \left[ \frac{\pi^4}{18} + \order{\ep} \right] \ , \\
\mathrm{Im}\ B(1,1,1,1,1,1,1,1,-1)
& = & \pi \mathcal{F}^3 \left[ \frac{17\pi^4}{720} + \order{\ep} \right] \ , \\
\mathrm{Im}\ E(1,0,1,1,1,1,1,1,0)
& = & \pi \mathcal{F}^3 \left[ \frac{\pi^4}{15} + \order{\ep} \right] \ , \\
\mathrm{Im}\ F(1,1,1,1,0,0,1,1,0)
& = & \pi \mathcal{F}^3 \left[ \left( -2 + \frac{\pi^2}{6} + \frac{13\zeta_3}{4} - \frac{\pi^2}{2}\ln2 \right) + \order{\ep} \right] \ , \\
\mathrm{Im}\ F(1,1,1,1,1,1,1,1,-1) & = & \pi \mathcal{F}^3 \left[
-\frac{\pi^4}{60} + \order{\ep} \right] \ .
\ea
 The other
integrals in~\cite{vanRitbergen:1999fi} are also relevant to our
calculation, however, since we require analytic results beyond the
leading order in $\ep$ for the remaining master integrals, we can
only use these four-loop integrals as a consistency check on our
own calculations.  The remaining eleven master integrals are: \ba
\label{M6}
\mathrm{Im} \left( \int \frac{[d^Dk_1][d^Dk_2][d^Dk_3]}{[k_1^2+1](k_1+k_2)^2k_2^2k_3^2(k_2+k_3+p)^2} \right)
& = & \pi \mathcal{F}^3 \left[ \frac{1}{2\ep} + \frac{15}{4} + \left( \frac{145}{8} - \frac{\pi^2}{2} \right) \ep \right. \nonumber \\
& & \qquad \left. + \left( \frac{1155}{16} - \frac{15\pi^2}{4} - 5\zeta_3 \right) \ep^2 + \order{\ep^3} \right] \ , \\
\mathrm{Im} \left( \int \frac{[d^Dk_1][d^Dk_2][d^Dk_3]}{k_1^2(k_1+p)^2k_2^2k_3^2[(k_1+k_2+k_3+p)^2+1]} \right)
& = & \pi \mathcal{F}^3 \left[ -\frac{1}{2\ep^2} - \frac{5}{2\ep} - \frac{17}{2} + \left( -\frac{49}{2} + 2\zeta_3 \right) \ep \right. \nonumber \\
& & \qquad \left. + \left( -\frac{129}{2} + 10\zeta_3 \right) \ep^2 + \order{\ep^3} \right] \ , \\
\mathrm{Im} \left( \int \frac{[d^Dk_1][d^Dk_2][d^Dk_3]}{(k_1+p)^2(k_1+k_2)^2(k_1+k_3)^2k_2^2(k_3^2+1)} \right)
& = & \pi \mathcal{F}^3 \left[ \frac{1}{2\ep} + \left( \frac{11}{2} - \frac{\pi^2}{6} \right) + \left( \frac{77}{2} - \frac{3\pi^2}{2} - 8\zeta_3 \right) \ep \right. \nonumber \\
& & \qquad \left. + \left( \frac{439}{2} - \frac{19\pi^2}{2} - 53\zeta_3 - \frac{3\pi^4}{10} \right) \ep^2 + \order{\ep^3} \right] \ .
\ea
\ba
\mathrm{Im}\ B(1,1,1,0,1,1,0,1,0)
& = & \pi \mathcal{F}^3 \left[ \frac{1}{2\ep^2} + \frac{7}{2\ep} + \left( \frac{33}{2} - \frac{\pi^2}{3} - 2\zeta_3 \right) + \left( \frac{131}{2} - \frac{7\pi^2}{3} - 10\zeta_3 - \frac{7\pi^4}{180} \right) \ep + \order{\ep^2} \right] \ , \\
\label{T6}
\mathrm{Im}\ B(1,1,1,1,1,2,0,0,0)
& = & \pi \mathcal{F}^3 \left[ \left( - \frac{5\zeta_3}{4} + \frac{\pi^2}{2}\ln2 \right) \right. \\
& & \qquad \left. + \left( - \frac{5\zeta_3}{2} + \frac{11\pi^4}{240} + \pi^2\ln2 + \frac{\pi^2}{3}\ln^22 - \frac{1}{3}\ln^42 - 8\mathrm{Li}_4(1/2) \right) \ep + \order{\ep^2} \right] \ , \nonumber \\
\label{T7}
\mathrm{Im}\ B(1,1,1,1,2,2,0,0,0)
& = & \pi \mathcal{F}^3 \left[ \frac{\pi^2}{12} + \left( \frac{\pi^2}{6} - \frac{3\zeta_3}{4} + \frac{\pi^2}{2}\ln2 \right) \ep \right. \\
& & \qquad \left. + \left( \frac{\pi^2}{3} - \frac{3\zeta_3}{2} + \frac{11\pi^4}{180} + \pi^2\ln2 + \frac{\pi^2}{3}\ln^22 - \frac{1}{3}\ln^42 - 8\mathrm{Li}_4(1/2) \right) \ep^2 + \order{\ep^3} \right] \ , \nonumber \\
\mathrm{Im}\ C(1,0,1,1,0,1,1,1,0)
& = & \pi \mathcal{F}^3 \left[ 1 + \left( 14 - 4\zeta_3 \right) \ep +
\left( 119 - \pi^2 - 28\zeta_3 - \frac{\pi^4}{3} \right) \ep^2 + \order{\ep^3} \right] \ , \\
\mathrm{Im}\ C(1,1,1,0,1,1,1,0,0)
& = & \pi \mathcal{F}^3 \left[ \frac{1}{2\ep^2} + \frac{5}{2\ep} + \left( \frac{15}{2}
 - \frac{\pi^2}{6} - 2\zeta_3 \right) + \left( \frac{29}{2} - \frac{\pi^2}{6} - 9\zeta_3
  + \frac{\pi^4}{60} \right) \ep + \order{\ep^2} \right] \ , \\
\label{M11}
\mathrm{Im}\ F(1,1,0,0,1,1,1,1,0)
& = & \pi \mathcal{F}^3 \left[ \left( -1 + \frac{\pi^2}{6} \right) +
\left( -14 + \frac{7\pi^2}{6} + 9\zeta_3 \right) \ep \right. \nonumber \\
& & \qquad \left. + \left( -119 + \frac{15\pi^2}{2} + 63\zeta_3 + \frac{41\pi^4}{180}
 \right) \ep^2 + \order{\ep^3} \right] \ , \\
\label{M17}
\mathrm{Im}\ F(1,1,0,1,0,1,1,1,0)
& = & \pi \mathcal{F}^3 \left[ \frac{1}{\ep} + \left( 11 - \frac{\pi^2}{6}
- \zeta_3 \right) + \left( 77 - \frac{13\pi^2}{6} - 13\zeta_3 - \frac{7\pi^4}{90} \right) \ep
 + \order{\ep^2} \right] \ , \\
\label{M16}
\mathrm{Im}\ G(1,1,1,0,1,1,1,0,0)
& = & \pi \mathcal{F}^3 \left[ \left( 1 - \frac{\pi^2}{6} + \zeta_3 \right)
+ \left( 14 - \frac{7\pi^2}{6} - 3\zeta_3 + \frac{\pi^4}{18} \right) \ep + \order{\ep^2} \right] \ .
\ea
The integrals in Eqs.~(\ref{T6}) and~(\ref{T7}) contain the fourth-order Polylogarithm
$\mathrm{Li}_4(1/2)=0.517479\ldots$, along with terms with higher powers of $\ln2$.
 None of these constants appear in our final results, which suggests that there might
 exist a basis of master integrals in which these constants do not arise in any of the
 terms at the orders in $\ep$ which contribute to our final results.

\section{Evaluation of Master Integrals}
The eleven master integrals in Eqs.~(\ref{M6}) through~(\ref{M16})
were evaluated on a case-by-case basis by integrating the phase
space of massless propagators severed by cuts.  In this section,
we will illustrate the procedure with a few examples. We will
compute master integrals having two-, three-, and four-particle
cuts, respectively.

The general procedure is as follows.  For any particular integral,
we can use the optical theorem to obtain its imaginary part via
\be \label{ImM} \mathrm{Im}\,\mathcal{M} = \frac{1}{2}
\sum_{\mathrm{cuts}} \int d\Pi_{\mathrm{cut}} \ \mathcal{M}_1
\mathcal{M}_2^{\ast} \ , \ee where $\mathcal{M}_1$ and
$\mathcal{M}_2$ are the scalar amplitudes arising from the uncut propagators
on the two sides of the cut, the cut propagators are assumed to be
on mass shell, and $\int d\Pi_{\mathrm{cut}}$ is the integral over
the phase space of the cut propagators.  Working in dimensional
regularization, the phase space integral involves a factor of \be
\frac{d^{D-1}\mathbf{q}}{(2\pi)^{D-1} 2E} \ee for every particle
involved in a given cut, along with a $D$-dimensional
$\delta$-function to conserve overall four-momentum.  Much of the
phase space integration can be done for a general case without
assuming anything about the amplitudes $\mathcal{M}_1$ and
$\mathcal{M}_2$, so that the phase space integrals relevant to an
$n$-particle cut can be reduced to simpler integrals involving a
small number of scalar parameters.  We shall see this explicitly
in the following three examples, each with a different number of
cut particle lines.

\subsection{Example 1: two-particle cut}
\begin{figure}[htb]
\hspace*{0mm}\psfig{figure=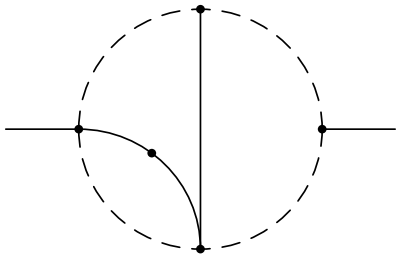,width=52mm} \caption{Sketch of
the master integral $B(1,1,1,1,1,2,0,0,0)$.  The imaginary part of
this integral can be obtained from a two-particle cut through the
massless lines on the right side of the diagram.  Solid and dashed lines depict
massive and massless propagators, respectively.} \label{fig:T6}
\end{figure}

Our first example will be the integral of Eq.~(\ref{T6}).  As can
be seen in the sketch of Fig.~\ref{fig:T6}, there is a single cut
(through the two rightmost propagators) that can be made through
massless lines in the diagram.
The two-body phase space integral for massless particles is
especially simple, since the two particles must emerge
back-to-back while partitioning the total energy equally.
Defining the general phase-space factor by
\be
\mathbf{P_n} =
 \int\; \prod_{i=1}^n {{{\rm d}^{D-1}\mathbf{p}_i\over (2\pi)^{D-1}2E_i}}
 \; (2\pi)^D \delta^D\left(p-\sum_{j=1}^n p_j\right) \ ,
\ee
an explicit calculation leads to the two-body phase space factor
\be
\label{P2}
\mathbf{P_2} = \frac{\mathcal{F}\ \ 2^{2\ep}
\pi^{3/2}}{\Gamma(1+\ep) \Gamma(3/2-\ep)} \ ,
\ee
where, as
before, $\mathcal{F}$ denotes the loop factor $\Gamma(1+\ep)/(4\pi)^{D/2}$.
There are no remaining integrals in
this expression, and so we have
\be \label{P2L1} \mathrm{Im}\
B(1,1,1,1,1,2,0,0,0) = \frac{\mathbf{P_2}}{2}\ L_1 \ ,
\ee
where
$L_1$ is the factor arising from the two-loop integral for the
three-point function to the left of the cut in Fig.~\ref{fig:T6}.
Explicitly,
\be L_1 = \int
\frac{[d^Dk][d^Dl]}{(k+p)^2[(k+p-q)^2+1][(k+l)^2+1]^2l^2} \ ,
\ee
where $k$ and $l$ are the loop momenta, $p$ is the incoming
momentum from the left (with the onshell condition $p^2=-1$), $q$
emerges onto the massless line on the upper-right ($q^2=0$), and
$(p-q)$ emerges onto the massless line on the lower-right (such
that $(p-q)^2=0$).  Applying the Feynman parameters $x$, $y$, and
$z$ in three consecutive stages, we can evaluate the $l$- and
$k$-integrals so that we are left with the parametric integral
\be
L_1 = \mathcal{F}^2 \ \frac{\Gamma(1+2\ep)}{\Gamma^2(1+\ep)}
\int_0^1 \!\! dx \int_0^1 \!\! dy \int_0^1 \!\! dz \
\frac{x^{\ep}(1-x)^{-\ep}z(1-z)^{\ep}}{\left[ z^2xy + zx(1-2y) +
1-z \right]^{1+2\ep}} \ .
\ee
 The $y$-integral can be solved
explicitly, but the remaining $x$- and $z$-integrals are more
difficult.  The integral is finite (as a result of our use of a
second factor of one of the massive lines) and thus we can expand
the integrand in $\ep$ using expressions like \be x^{\ep} = 1 +
\ep \ln x + \frac{\ep^2}{2} \ln^2 x + \order{\ep^3} \ , \ee and
then evaluate the integrals order by order in $\ep$.  The leading
($\ep^0$) term is relatively simple, but the presence of products
of logarithms in the subsequent terms quickly escalates the level
of difficulty.  The FORM packages Summer~\cite{Vermaseren:1998uu}
and Harmpol~\cite{Remiddi:1999ew} have proven to be very useful
for evaluating some of these integrals; some of the related formulas
are conveniently tabulated in \cite{blumlein}. For the first two
terms of $L_1$, we obtain
\be L_1 = \mathcal{F}^2 \left[ \left( -
\frac{5\zeta_3}{4} + \frac{\pi^2}{2}\ln2 \right) + \left( -
\frac{5\zeta_3}{2} + \frac{11\pi^4}{240} + \pi^2\ln2 +
\frac{\pi^2}{3}\ln^22 - \frac{1}{3}\ln^42 - 8\mathrm{Li}_4(1/2)
\right) \ep + \order{\ep^2} \right] \ ,
\ee
 and, in conjunction
with Eqs.~(\ref{P2}) and~(\ref{P2L1}), thereby establish the
result in Eq.~(\ref{T6}).

\subsection{Example 2: three-particle cut}
\begin{figure}[htb]
\hspace*{0mm}\psfig{figure=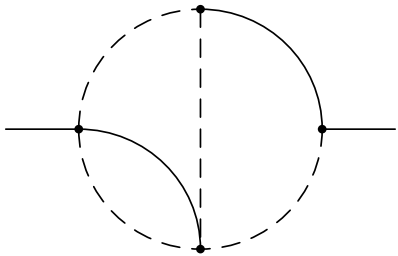,width=52mm} \caption{Sketch of
the master integral $F(1,1,0,1,0,1,1,1,0)$.  The imaginary part of
this integral can be obtained from a three-particle cut through
the massless lines from the upper left side of the diagram towards
the lower right.} \label{fig:M17}
\end{figure}

An example of a master integral requiring a three-particle cut is
Eq.~(\ref{M17}).  This integral is sketched in Fig.~\ref{fig:M17},
where it is seen that a diagonal cut through the massless lines
leaves a one-loop amplitude to the left of the cut and a single
massive propagator to the right.
The three massless particles that we cut through can be put on
shell with a range of relative kinematic configurations, meaning
that the three-body phase space integral can only be partially
solved in advance.  The remainder of the integral will be over two
kinematic parameters (such as the energies of two of the
particles) upon which the amplitudes $\mathcal{M}_1$ and
$\mathcal{M}_2$ might also depend.  A very convenient formulation
of the three-body phase space involves the use of the relativistic
invariants
\be
u=(q_1+q_3)^2 \qquad \mbox{and} \qquad
z=(q_2+q_3)^2
\ee
as the integration variables.  This leads to the
expression
\be
\mathbf{P_3} = \frac{(\mathbf{P_2})^2}{2\pi B_{11}}
\int_0^1 \!\! dz \int_0^{1-z} \!\! du \ u^{-\ep} \ z^{-\ep} \
(1-z-u)^{-\ep} \ , \label{P3}
\ee
where $B_{11}=B(1-\ep,1-\ep)$.
Returning to the master integral in Fig.~\ref{fig:M17}, we observe
that the massive propagator ($m=1$) in the upper right part of the
diagram splits directly into two onshell massless particles, and
thus the momentum $q$ flowing through this propagator is related
to one of our invariants:
\be
(q^2+1)^{-1} \longrightarrow
(1-z)^{-1} \ .
\ee
 In addition, we must include a factor in
the phase space integral resulting from the one-loop integral in
the lower left portion of Fig.~\ref{fig:M17}.  The momentum $l$
flowing through this diagram is related to the other invariant via
$l^2=-u$, and the loop integral can quickly be evaluated with a
Feynman parameter:
\be \int \frac{[d^Dk]}{[(k+l)^2+1]k^2} =
\frac{\mathcal{F}}{\ep} \int_0^1 \!\! dx \ (1-x)^{-\ep}
(1-xu)^{-\ep} \ .
\ee
Substituting this expression into
Eq.~(\ref{P3}), we have
\be
\mathrm{Im}\ F(1,1,0,1,0,1,1,1,0) =
\frac{\mathcal{F} \ (\mathbf{P_2})^2}{4\pi \ep B_{11}} J \ ,
\label{M17p1}
\ee
where $J$ denotes the 3-parameter integral:
\be
J = \int_0^1 \!\! dx \int_0^1 \!\! dz \int_0^{1-z} \!\! du \
\frac{(1-x)^{-\ep}\ u^{-\ep} \ z^{-\ep} \ (1-z-u)^{-\ep}\
(1-xu)^{-\ep}}{(1-z)} \ .
\ee
As in the previous example, this
integral is finite, and hence we can safely expand the integrand
in $\ep$ and evaluate it term by term.  The result,
\be J = 1 +
\left( 9 - \frac{\pi^2}{6} - \zeta_3 \right) \ep + \left( 55 -
\frac{4\pi^2}{3} - 11\zeta_3 - \frac{7\pi^4}{90} \right) \ep^2 +
\order{\ep^3} \ ,
\ee
once folded in with the other factors in
Eq.~(\ref{M17p1}), leads to the expression of Eq.~(\ref{M17}).

\subsection{Example 3: four-particle cut}
\begin{figure}[htb]
\hspace*{0mm}\psfig{figure=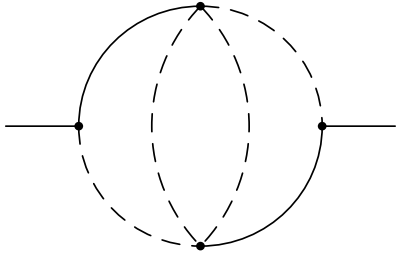,width=52mm} \caption{Sketch of
the master integral $F(1,1,0,0,1,1,1,1,0)$.  The imaginary part of
this integral can be obtained from a four-particle cut through the
massless lines.} \label{fig:M11}
\end{figure}

Our final example illustrates how a four-particle cut can be used
to obtain Eq.~(\ref{M11}).  This integral is sketched in
Fig.~\ref{fig:M11}.
On each side of the cut we have a massive propagator.  As might be
expected when cutting through four lines in a planar three-loop diagram,
there are no loop integrals remaining.  The most general
expression for a four-body phase space integral is extremely
complicated, but it happens that we can obtain a simpler
expression suitable for our purposes by modifying the three-body
result.  Specifically, the generalization of the three-body phase
space integral in Eq.~(\ref{P3}) from the case of three massless
particles to the case of two massless particles and a particle of
squared-mass $y=q_2^2$ (such that $0\leq y \leq 1$) leads to
\be
\mathbf{P_{3m}} = \frac{(\mathbf{P_2})^2}{2\pi B_{11}} \int_0^1 \!\!
dz \int_0^{1-z} \!\! du \ u^{-\ep} \ z^{-\ep} \ \left(
\frac{(1-z)(z-y)}{z} -u \right)^{-\ep} \ .
\label{P3m}
\ee
When
$y=0$, this reduces to Eq.~(\ref{P3}), but if we interpret $y$ as
the relativistic invariant of a pair of massless particles in a
four-body phase space,
\be y=(q_2+q_4)^2 \ ,
\ee
we obtain a
four-body phase space integral
\be
\mathbf{P_4} =
\frac{(\mathbf{P_2})^3}{4\pi^2 B_{11}} \int_0^1 \!\! dz
\int_0^{1-z} \!\! du \int_0^{z(1-z-u)/(1-z)} \!\! dy \ u^{-\ep} \
z^{-\ep} \ y^{-\ep} \ \left( \frac{(1-z)(z-y)}{z} -u
\right)^{-\ep} \label{P4u}
\ee
 that is applicable to certain
integrals where two of the massless particles form a subloop.  In
order to evaluate the integral in Fig.~\ref{fig:M11}, we will need
to replace the invariant $u=(q_1+q_3)^2$ with $v=(q_1+q_2)^2$
since the massless loop corresponds to a massive $q_2$ and thus
the massive propagators beside the cut carry momenta $(q_1+q_2)$
and $(q_2+q_3)$.  With \be u+v+z=1+y \ , \ee Eq.~(\ref{P4u}) can
be rewritten as \be \mathbf{P_4} = \frac{(\mathbf{P_2})^3}{4\pi^2
B_{11}} \int_0^1 \!\! dy \int_y^1 \!\! dz \int_{y/z}^{1+y-z} \!\!
dv \ y^{-\ep} \ z^{-\ep} \ (1+y-z-v)^{-\ep} \ \left( v -
\frac{y}{z} \right)^{-\ep} \ . \label{P4v} \ee The integral we
need to evaluate has the additional propagator factors
$(1-v)^{-1}$ and $(1-z)^{-1}$, and as before, can be expanded in
$\ep$ and evaluated term by term, thereby leading to the result of
Eq.~(\ref{M11}).

\section{Results}
\label{results}
The QCD corrections to the decay width for $t\rightarrow bW$ can
be written as
\be
\Gamma(t\rightarrow bW) = \Gamma_0 \left[ X_0 +
\frac{\alpha_s}{\pi} X_1 + \left( \frac{\alpha_s}{\pi} \right)^2
X_2 + \order{\alpha_s^3} \right] \ , \label{GammaX} \ee where
\vspace{-7mm} \be \Gamma_0 = \frac{G_F \left| V_{tb} \right|^2
m_t^3}{8\sqrt{2}\pi} \ ,
\label{G0}
\ee
is the tree-level decay
width in the limit that $m_W=0$.
The results for $X_0$ and $X_1$
are known analytically \cite{jk2,Jezabek:1993wk}.

The two-loop contribution with
which we are presently concerned, $X_2$, has been estimated numerically
\cite{Czarnecki:1998qc,Chetyrkin:1999ju}.  Detailed comparisons of our analytical
results with those studies is given in \cite{Blokland:2004ye}.
We divide $X_2$
into four gauge-invariant color structures,
\be
X_2 = C_F
\left( T_R N_L X_L + T_R N_H X_H + C_F X_A + C_A X_{NA} \right) \
. \label{X2pieces}
\ee
Here we use $SU(N)$ factors (with $N=3$) $C_F=(N^2-1)/(2N)$, $C_A=N$, and $T_R=1/2$.
$N_L$ denotes the number of massless quark species and $N_H$ that of quarks with
the mass equal to that of the decaying quark.
Defining the parameter
$\omega=m_W^2/m_t^2$, our results for the four classes of
contributions in Eq.~(\ref{X2pieces}) are:
\ba
X_L & = & \left[
-\frac{4}{9} + \frac{23\pi^2}{108} + \zeta_3 \right] + \omega
\left[ -\frac{19}{6} + \frac{2\pi^2}{9} \right]
+ \omega^2 \left[ \frac{745}{72} - \frac{31\pi^2}{36} - 3\zeta_3 - \frac{7}{4}\ln\omega \right] \nonumber \\
& & + \omega^3 \left[ -\frac{5839}{648} + \frac{7\pi^2}{27} + 2\zeta_3 + \frac{5}{3}\ln\omega \right]
+ \omega^4 \left[ \frac{4253}{8640} + \frac{\pi^2}{4} - \frac{17}{144}\ln\omega \right] \nonumber \\
& & + \omega^5 \left[ -\frac{689}{27000} + \frac{\pi^2}{15} - \frac{7}{900}\ln\omega \right]
+ \omega^6 \left[ -\frac{13187}{181440} + \frac{\pi^2}{36} + \frac{1}{48}\ln\omega \right] \nonumber \\
& & + \omega^7 \left[ -\frac{2282381}{37044000} + \frac{\pi^2}{70} +
\frac{2263}{88200}\ln\omega \right] + \order{\omega^8} \ ,
\label{XL}
\ea
\ba
X_H & = & \left[ \frac{12991}{1296} - \frac{53\pi^2}{54} - \frac{\zeta_3}{3} \right]
+ \omega \left[ -\frac{35}{108} - \frac{4\pi^2}{9} + 4\zeta_3 \right]
+ \omega^2 \left[ - \frac{6377}{432} + \frac{25\pi^2}{18} + \zeta_3 \right] \nonumber \\
& & + \omega^3 \left[ \frac{319}{27} - \frac{31\pi^2}{27} - \frac{2\zeta_3}{3} \right]
+ \omega^4 \left[ \frac{76873}{8640} - \frac{8\pi^2}{9} \right]
+ \omega^5 \left[ \frac{237107}{27000} - \frac{8\pi^2}{9} \right] + \order{\omega^6} \ ,
\label{XH}
\ea
\ba
X_A & = & \left[ 5 - \frac{119\pi^2}{48} - \frac{53\zeta_3}{8} - \frac{11\pi^4}{720} + \frac{19}{4}\pi^2\ln2 \right]
+ \omega \left[ -\frac{73}{8} + \frac{41\pi^2}{8} - \frac{41\pi^4}{90} \right] \nonumber \\
& & + \omega^2 \left[ -\frac{7537}{288} + \frac{523\pi^2}{96} + \frac{295\zeta_3}{32} - \frac{191\pi^4}{720} - \frac{27}{16}\pi^2\ln2 + \left( \frac{115}{48} - \frac{5\pi^2}{16} \right) \ln\omega \right] \nonumber \\
& & + \omega^3 \left[ \frac{16499}{864} - \frac{407\pi^2}{216} - \frac{7\zeta_3}{2} + \frac{7\pi^4}{120} - \pi^2\ln2 + \left( -\frac{367}{144} + \frac{5\pi^2}{9} \right) \ln\omega \right] \nonumber \\
& & + \omega^4 \left[ -\frac{1586479}{259200} + \frac{2951\pi^2}{6912} + \frac{9\zeta_3}{2} + \left( \frac{31979}{17280} - \frac{\pi^2}{16} \right) \ln\omega \right] \nonumber \\
& & + \omega^5 \left[ -\frac{11808733}{6480000} + \frac{37\pi^2}{2400} + \frac{6\zeta_3}{5} + \left( \frac{13589}{27000} - \frac{\pi^2}{60} \right) \ln\omega \right]
+ \order{\omega^6} \ ,
\label{XA}
\ea
\ba
X_{NA} & = & \left[ \frac{521}{576} + \frac{505\pi^2}{864} + \frac{9\zeta_3}{16} + \frac{11\pi^4}{1440} - \frac{19}{8}\pi^2\ln2 \right]
+ \omega \left[ \frac{91}{48} + \frac{329\pi^2}{144} - \frac{13\pi^4}{60} \right] \nonumber \\
& & + \omega^2 \left[ -\frac{12169}{576} + \frac{2171\pi^2}{576} + \frac{377\zeta_3}{64} - \frac{77\pi^4}{288} + \frac{27}{32}\pi^2\ln2 + \left( \frac{73}{16} - \frac{3\pi^2}{32} \right) \ln\omega \right] \nonumber \\
& & + \omega^3 \left[ \frac{13685}{864} - \frac{47\pi^2}{72} - \frac{19\zeta_3}{4} + \frac{43\pi^4}{720} + \frac{1}{2}\pi^2\ln2 + \left( -\frac{1121}{432} - \frac{\pi^2}{6} \right) \ln\omega \right] \nonumber \\
& & + \omega^4 \left[ -\frac{420749}{103680} - \frac{3263\pi^2}{13824} - \frac{9\zeta_3}{8} + \left( \frac{11941}{6912} - \frac{3\pi^2}{32} \right) \ln\omega \right] \nonumber \\
& & + \omega^5 \left[ -\frac{4868261}{12960000} - \frac{557\pi^2}{4800} - \frac{3\zeta_3}{10} + \left( \frac{153397}{216000} - \frac{\pi^2}{40} \right) \ln\omega \right]
+ \order{\omega^6} \ .
\label{XNA}
\ea

\begin{figure}[htb]
\hspace*{0mm}\psfig{figure=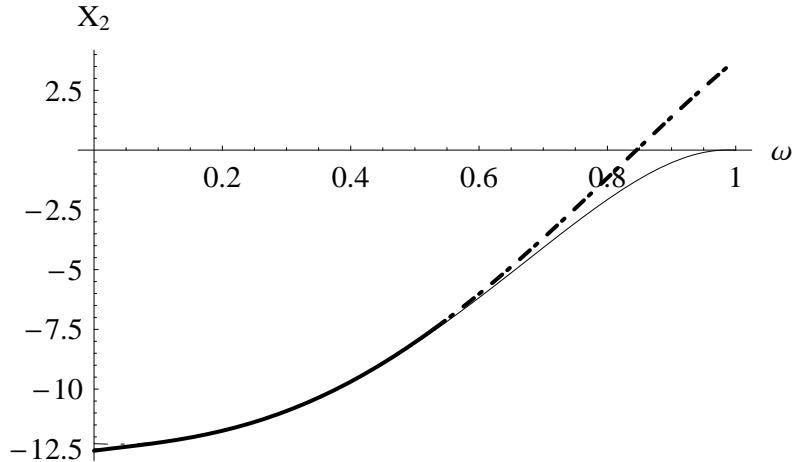,width=105mm}
\caption{Matching
of expansions around $\omega = 0$ (thick line) and $\omega = 1$
(thin line). The solid line denotes the resulting correction to the decay width
valid in the full range of $\omega$.}
\label{fig:expansion}
\end{figure}

As discussed in~\cite{Blokland:2004ye}, we obtain the two-loop QCD
correction to the top quark decay width by substituting
$\omega\simeq 0.213$ into these expressions.  There are already
sufficiently many terms so as to render the theoretical
uncertainty 20 times smaller than the experimental uncertainty
induced by the mass measurements of the top quark.  Furthermore, these
expressions can be smoothly matched to the corresponding
expressions~\cite{Czarnecki:2001cz} for the two-loop QCD
corrections to the semileptonic decay $b\to u\ell \nu$ in the zero
recoil limit. The result of such a matching procedure is depicted
in Fig.~\ref{fig:expansion}.

\section{Conclusions}
We have presented an analytical approach to
next-to-next-to-leading order studies of heavy-to-light decays.
We have used top quark decay as an illustrative example here; in
\cite{Blokland:2004ye} we also studied the distribution of the
lepton invariant mass in the semileptonic $b\to u$ decay and have
reproduced the $\order{\alpha^2}$ corrections to the muon
lifetime.  A number of other phenomenologically interesting
studies become possible with the results presented here.  For
example, corrections to radiative $b\to s\gamma$ decays and to
$B_{d,s}$ meson mixing require matrix elements of very similar if
not identical kinematic structure.  (For the mixing, one can expand in the
four-momentum carried by the light quark and obtain propagator-like
diagrams studied in the present paper.)

We employed two approaches to reducing Feynman integrals to a
basis set of primitive ones. In both we use integration-by-parts
recurrence relations that we either solve completely, or for a
targeted set of needed integrals.

The most appealing feature of the targeted approach is that the core of
the program is  universal for all topologies.  Unlike the
complete solution method, where each new topology necessitates a by-hand
construction of a new solution algorithm, the targeted approach
needs only a minimal specification of the topology under
consideration.  As a result, once the program core is written and
tested, very little programming is needed to incorporate new
topologies, thereby saving time, effort, and minimizing the
likelihood of programming bugs.

Another major difference between the two approaches is the use of
subtopologies.  In the traditional method, the removal of a line
will often facilitate a mapping of a loop integral from a more
complex topology onto a simpler topology, for which a new set of
identities is used.  The drawback to this intuitively reasonable
process is a proliferation in the number of topologies to be
programmed.  For example, the 9 master topologies that we
encounter in this work (discussed in  section \ref{sec:topo})
require approximately 40 additional subtopologies.  The
targeted approach, on the other hand, only needs to deal with
the master topologies, as it is able to solve a given topology
completely using only one set of identities.  At the three-loop
level, it would be very difficult for a human to program an
algorithm in this way.

A key feature of the targeted approach is the generation of
large tables containing all possible integrals in a topology with
a predetermined range of powers.  Once this table is generated,
it is accessed by the programs responsible for assembling a
particular physical calculation.  In the complete method,
conversely, the algorithms are typically applied in real-time, and
the generation of dynamic tables will likely be used in the near
future to improve the performance of this approach.  Given the
large range of powers which can arise when performing a
calculation using asymptotic expansions, these tables and intermediate
expressions can become quite large.

In summary, the targeted approach appears to be superior by virtue
of saving human time.  The burden is shifted to the computer and
thus an efficient implementation of the algorithm is essential. We
developed the project-specific software based on the
BEAR package \cite{bear}.  The targeted approach has also recently
been implemented in MAPLE \cite{Anastasiou:2004vj}.

The availability of the targeted approach greatly simplifies
multiloop calculations. The remaining bottleneck involves the
evaluation of master integrals. Further  progress in perturbative
calculations may depend on a radically new approach addressing this issue.

\appendix*
\section{}
As described in Section \ref{results}, expansion around $\omega=0$ is
given by the functions (\ref{XL})--(\ref{XNA}). The complementary expansion
around $\omega=1$ has been studied in \cite{Czarnecki:2001cz}.
Below, the corresponding formulae are given for completeness:
\ba
X_L ( \delta \simeq 0 ) & = & \delta^2 \left[ - \frac{117}{8} + \frac{41}{18} \pi^2 + 6 \zeta_3 +
                               \left( \frac{39}{4} - \frac{4}{3} \pi^2 \right) \ln \delta
                                - \frac{3}{2} \ln^2 \delta \right] \nonumber \\
& & + \delta^3 \left[ \frac{797}{108} - \frac{53}{27} \pi^2 - 4
\zeta_3 +
                            \left( - \frac{31}{18} + \frac{8}{9} \pi^2 \right) \ln \delta
                          - \frac{1}{3} \ln^2 \delta \right] \nonumber \\
& & + \delta^4 \left[ \frac{1289}{432} + \frac{1}{18} \pi^2 -
\frac{37}{18} \ln \delta +
     \frac{1}{6} \ln^2 \delta \right] + \delta^5 \left[ \frac{1817}{3600} - \frac{3}{5} \ln \delta \right]
+ \order{\delta^6} \ ,
\\
X_H ( \delta \simeq 0 ) & = & \delta^2 \left[ \frac{133}{16} - \frac{5}{6} \pi^2 \right] +
    \delta^3 \left[ -\frac{797}{216} + \frac{1}{3} \pi^2 \right] +
    \delta^4 \left[ \frac{2473}{5400} - \frac{1}{36} \pi^2 - \frac{1}{10} \ln \delta \right]
     \nonumber \\
& & + \delta^5 \left[ \frac{1747}{5400} - \frac{1}{30} \pi^2 -
\frac{1}{15} \ln \delta \right] + \order{\delta^6} \ ,
\\
 X_A
( \delta \simeq 0 ) & = & \delta^2 \left[ \frac{523}{32} -
\frac{71}{12} \pi^2 + \frac{7}{2} \pi^2 \ln 2
                              + \frac{8}{15} \pi^4 - \frac{39}{4} \zeta_3 -
                              \left(\frac{147}{16} - 2 \pi^2 \right) \ln \delta +
                              \frac{27}{8} \ln^2 \delta \right] \nonumber \\
& & + \delta^3 \left[ \frac{1363}{144} - \frac{217}{216} \pi^2 -
\frac{1}{3} \pi^2 \ln 2 - \frac{32}{90} \pi^4 +
                          \left( - \frac{287}{24} + \frac{5}{9} \pi^2 \right) \ln \delta  +
                   \frac{15}{4} \ln^2 \delta \right] \nonumber \\
& & + \delta^4 \left[ \frac{1537}{288} + \frac{937}{864} \pi^2 -
\frac{5}{12} \pi^2 \ln 2 -
                             \left(\frac{67}{12} + \frac{17}{36} \pi^2 \right) \ln \delta  +
                          \frac{17}{8} \ln^2 \delta  \right] \nonumber \\
& &  +\delta^5 \left[ \frac{43609}{43200} + \frac{7609}{21600}
\pi^2 - \frac{1}{4} \zeta_3 -
                          \frac{1}{10} \pi^2 \ln 2 -
                          \left( \frac{269}{90} + \frac{8}{90} \pi^2 \right) \ln \delta +
                          \frac{11}{6} \ln^2 \delta \right]
                          + \order{\delta^6} \ ,
\\ X_{NA} ( \delta \simeq 0) & = & \delta^2 \left[
\frac{1103}{32} - \frac{881}{144} \pi^2 - \frac{7}{4} \pi^2 \ln 2
                                 + \frac{13}{60} \pi^4 - \frac{129}{8} \zeta_3 +
                                 \left( - \frac{423}{16} + \frac{47}{12} \pi^2 \right) \ln \delta
                                  +
                                 \frac{33}{8} \ln^2 \delta \right] \nonumber \\
& & + \delta^3 \left[ -\frac{623}{54} + \frac{247}{48} \pi^2 +
\frac{1}{6} \pi^2 \ln 2
                             - \frac{13}{90} \pi^4 + 12 \zeta_3 +
                             \left( \frac{155}{72} - 2 \pi^2 \right) \ln \delta  +
                             \frac{2}{3} \ln^2 \delta \right] \nonumber \\
& & +\delta^4 \left[ - \frac{18319}{1728} - \frac{19}{572} \pi^2 +
\frac{5}{24} \pi^2 \ln 2
                            - \frac{1}{8} \zeta_3 +
                             \left( \frac{1877}{288} - \frac{1}{24} \pi^2 \right) \ln \delta  -
                            \frac{17}{24}  \ln^2 \delta \right] \nonumber \\
& & + \delta^5 \left[ - \frac{52379}{43200} - \frac{5041}{43200}\pi^2 +
\frac{1}{20} \pi^2 \ln 2 -
                              \frac{1}{40} \zeta_3 +
                              \left( \frac{139}{90} + \frac{1}{90} \pi^2 \right) \ln \delta  -
                            \frac{11}{48} \ln^2 \delta \right]
                            + \order{\delta^6} \ ,
\ea
where $\delta = 1 - \omega$.
These two sets of expressions can be smoothly matched in the
intermediate region to produce functions valid in the full range of
$\omega$. For practical purposes it is convenient to replace this
complicated combination of exact expansions by a collection of polynomial
fits with numerical coefficients. We found that the following simple functions
\begin{eqnarray}
X_L & \simeq & 2.854 - 0.665 \, \omega - 0.109 \, \omega^2 - 8.572
\, \omega^3 + 5.561 \, \omega^4 + 0.931 \, \omega^5 \ ,
\\
X_H & \simeq & -0.063615 + 0.098146 \, \omega + 0.144642 \,
\omega^2 - 0.307331 \, \omega^3 + 0.107417 \, \omega^4 + 0.020707
\, \omega^5 \ ,
\\
X_A & \simeq & 3.575 - 2.867 \, \omega + 2.241 \, \omega^2 -
12.027 \, \omega^3 + 11.564 \, \omega^4 - 2.489 \, \omega^5 \ ,
\\
X_{NA} & \simeq & -8.151 + 2.990 \, \omega - 3.537 \, \omega^2 +
36.561 \, \omega^3 - 42.275 \, \omega^4 + 23.899 \, \omega^5 -9.494\,\omega^6\ ,
\end{eqnarray}
give a very good approximation. For any value of $\omega$ they
deviate from the exact value by less than $0.01$. The second order
contribution to the decay rate for $N_L = 5$ and $N_H = 1$ can be
written in the approximate form
\begin{equation}
X_2 \simeq -16.729 + 3.064 \, \omega + 4.481 \, \omega^2 + 35.191
\, \omega^3 - 10.336 \, \omega^4 - 15.655 \, \omega^5,
\end{equation}
where the error introduced by the fit is smaller than the
uncertainty of the result due to, for example, higher-order QCD
corrections.

\emph{Acknowledgements:} We thank Kirill Melnikov and Sven Moch
for helpful discussions and Johannes Bl\"umlein for sharing
Ref.~\cite{blumlein} with us.

This research was supported by the Science and Engineering
Research Canada, Alberta Ingenuity, and by the Collaborative
Linkage Grant PST.CLG.977761 from the NATO Science Programme.

FT thanks Andrzej Czarnecki for the kind hospitality at the
University of Alberta.

Part of this work was carried out while AC and M\'S were
visiting the Kavli Institute for Theoretical Physics in Santa Barbara,
with partial support by the National Science Foundation under Grant No. PHY99-07949.


\begin{thebibliography}{10}
\bibitem{targeted}
In common mathematical parlance this solution could be called ``brute force''.
  Since this does not do justice to the intellectual effort required for an
  efficient implementation of this approach, we use the word ``targeted''.
\bibitem{csc}
Our computations are performed with the facilities of the Centre for Symbolic
  Computation at the University of Alberta.
\end{thebibliography}

\end{document}